\documentstyle[prl,aps,aps12,psfig,times,floats]{revtex}
\input epsf
 
\newcommand{\ee}{\end{equation}}
\newcommand{\be}{\begin{equation}}

\newcommand{\bef}{\begin{figure}}
\newcommand{\eef}{\end{figure}}
\newcommand{\hmp}{\rm h^{-1}Mpc}
\title {Angular Projections of Fractal Sets}
\author {Ruth Durrer 
\and
Jean-Pierre Eckmann}
\address{D\'ept.~de Physique Th\'eorique, Universit\'e de Gen\`eve, 
24, Quai E. Ansermet, CH-1211 Gen\`eve, Switzerland}
\author{Francesco Sylos Labini${}^1$
\and
Marco Montuori${}^1$
\and
Luciano Pietronero${}^{1,2}$
}
\address{${}^1$Dipartimento di Fisica, Universit\'a di Roma La Sapienza and
INFM Sezione Roma1,
Piazzale A.~Moro~2, 00185 Roma, Italy}
\address{${}^2$ITCP Trieste, Italy}
\begin{document}
\maketitle

\begin{abstract}
We discuss various questions which arise when one considers the
central projection of three dimensional fractal sets (galaxy catalogs)
onto the celestial globe.
The issues are related to how fractal such projections look.
First we show that the lacunarity in the projection can be arbitrarily
small. Further characteristics of the projected set---in 
particular scaling---depend
sensitively on how the apparent sizes of galaxies are taken into
account. Finally, we discuss the influence of opacity of galaxies.
Combining these ideas, seemingly contradictory statements about
lacunarity and apparent projections can
be reconciled. \vspace{1cm}
\end{abstract}

The distribution of galaxies in the universe poses some intriguing puzzles.
A group of physicists led by one of the authors of the present letter
(L.P.), found by statistical analysis of all publically available
 three dimensional galaxy catalogs, that galaxies are distributed 
fractally with a fractal dimension of about 2 up to the largest 
distances available, namely $(50 - 100) \hmp$ \cite{Pietronero}. 
On the other hand, there are strong arguments which favor homogeneity
{\em i.e.,} a dimension of 3, in a range of up to $3000 \hmp$, which
is the Hubble scale \cite{Peebles,Davis}. At distances of the order of
this limit, largely unknown evolutionary
effects probably influence the galaxy distribution, which complicates
the analysis.

In this Letter, we study some of the problems related to the
observations at ``large'' distances up to the Hubble limit.

General theoretical arguments lead to a prediction of a
{\em homogeneous} universe, which has dimension 3, while the analysis
performed in \cite{Pietronero} indicate a dimension of about 2, as far as 
three-dimensional data are available. 
What is even more intriguing is that the {\em projection} of the available
data onto the celestial globe shows {\em no} apparent lacunarity and
this seems in flagrant contradiction with the dimension of 2 found in 
\cite{Pietronero}, when one
considers the data as a point set in 3 dimensions.
In a widely circulated letter, Peebles has asked whether the mathematics
of fractals allows for such a phenomenon.

Our---perhaps surprising---result
is that if one takes finite size effects properly into account, {\em 
a set of dimension 2 in 3-dimensional space can have a non-fractal and
quite homogeneous projection
onto the celestial globe}.
We will explain this result below and put it into perspective with known
mathematical facts.
Depending on how finite size effects are taken into account, sometimes
the projection will be homogeneous and sometimes it will be fractal.
In particular, we will argue 
that {\em there is no contradiction between observing 
a fractal dimension 2 in 3-dimensional space and a uniform projection 
of equal size points onto the celestial globe}. 

Of course, 
this result leaves wide open the more fundamental
question of how a galaxy distribution of dimension
of 2 can be reconciled with the homogeneity of the universe predicted by the 
currently accepted cosmological models. We have no answer to
this puzzle. In this letter we just show that
absence of a unique
interpretation of the measurements---to have more
confidence in the data. 

In order to explain the ``projection paradox'' and to put it into 
context, we first explain some mathematical aspects 
of the problem. A set ${\bf A}$ in ${\bf R}^3$ is said to have 
Hausdorff dimension 
$d$ if it can be covered by sets $S_{i, \delta}$ of diameter less than
$\delta$ in such a way that the ``Hausdorff measure''
\begin{equation}
H^s({\bf A}) \,=\, \liminf_{\delta \rightarrow 0} \sum_i ({\rm
diam}S_{i, \delta})^s 
\end{equation}
is zero for $s > d$ and infinite for $s<d$. For $s=d$ 
one can have $H^d({\bf A}) = \infty$ or $H^d({\bf A}) < \infty$. 
Thus, if the Hausdorff dimension is $d$, then $H^d({\bf A})$ can be finite or
infinite.
When $d$ is an integer and
$H^d({\bf A})$ is finite, one can decompose $\bf A$ into a regular part
(consisting of piecewise rectifiable sets of dimension $d$, such as
lines or sheets) and a singular part (consisting of ``dust''), see
\cite{Falconer}, Section~6.2, \cite{Mattila}, Section~9. For
non-integer $d$ the regular part is absent.
Almost every projection of the regular part
onto a
$d$-dimensional plane is of positive measure,
while all projections of the singular part (which is the interesting
case for the study of galaxies) have measure 0, {\em i.e.,} they are very
small. Finally,
if $H^d({\bf A})$ is infinite, then a clever method,
called the ``Venetian blind construction'' shows that the projection
onto {\em any} subspace can be prescribed, and can be essentially
anything we want. For example, one can construct a set in 3-space
whose projection onto all two-dimensional planes is a ``sundial'' in
the sense that it shows the hour (minutes, and seconds) of the current
time in Roman numerals (\cite{Falconer}, Section~6.3).

Thus, in the light of a strict mathematical definition, the answer to 
Peebles' question is that the projection is of positive measure 
(and hence relatively smooth) {\em only} in the cases when 
$d > 2$  or when $d=2$ and either $H^2  = \infty$ or $\bf A$
has a regular part. In this last
case, the set $\bf A$ must contain rectifiable
``lines'' or ``sheets.'' We disregard this situation as physically
irrelevant because galaxies are considered as points or tiny disks 
in this analysis and thus form a ``dust-like,'' {\em i.e.,} irregular set.

In this Letter, we 
argue that this is not the whole story, because some finite size 
effects come into play in a subtle and beautiful way. 
To demonstrate this point more clearly, we first work with 1-dimensional
sets in a 2-dimensional space and then illustrate the extension 
to 2-dimensional sets in 3-space.

We say that a set (of galaxies) has an effective dimension 1 inside
a physical space of dimension 2 at length scale $r$ if the mean number 
$n(r) dr$ of galaxies between distance $r$ and $r + dr$ 
goes like $C dr$, where $C$ is a constant. Integrating from 0 to $r$
we find for the total number of galaxies, $N$:
\begin{equation}
N(r) \,= \,\int_0^r dr'\,n(r') \,\sim\, C r~,
\label{2}
\end{equation}
and this is characteristic of the distribution of a set of 
dimension 1.
(For a 2-dimensional set, the corresponding laws are $n(r) \sim C' r$, and
$N(r) \sim C'r^2/2$.)
Eq.~\ref{2} shows the well-known fact that if the dimension of
a set is not equal to the dimension of the space in which it is
embedded, then the number density goes down (in our case
like $1/r$). This is precisely the cosmologically
intriguing aspect when dealing with distributions
which are not of maximal dimension. 

We next study what happens when we
project such a set onto the celestial globe 
(the unit circle in the case of a 2-dimensional universe).
If the galaxies are considered just as a countable set of points, the
question
of the measure of their projection makes no sense, since it is equal
to 0 by definition. This is, however, not what one means by the
dimension of an experimentally measured set, where anyway only a
finite number of points occur \cite{EckmannRuelle}. In fact, dimension
is experimentally a notion which holds only over a certain range of scales.
Even taking that into account,
the mathematical theorems cited above tell us 
that the projected density has {\em zero} measure
when $H^1$ is finite and the set is singular.

Next, we analyze in more detail
various aspects which
come into play when one considers the projections of galaxies onto the
celestial globe.
The issues we discuss here are {\em lacunarity}, the r\^ole of taking
into account
{\em apparent sizes}, and the influence of {\em opacity}. We first
define these quantities. The lacunarity describes the sizes of voids
between the galaxies. These voids can be large in
a set of small dimension, but we shall show that they can be arbitrarily
small in the projection. The apparent size problem has to do with
whether we represent galaxies as points of equal size, or apparent
size.
The first representation will be called {\em pixel projection}
and the second {\em apparent size projection}.
Finally, opacity is related to the following
observational problem: We assume there is a limit to how close
together two galaxies can be observed, and this means that little
opaque disks are drawn around each observed galaxy.

\begin{figure}[h]
\epsfxsize=6cm
\epsffile{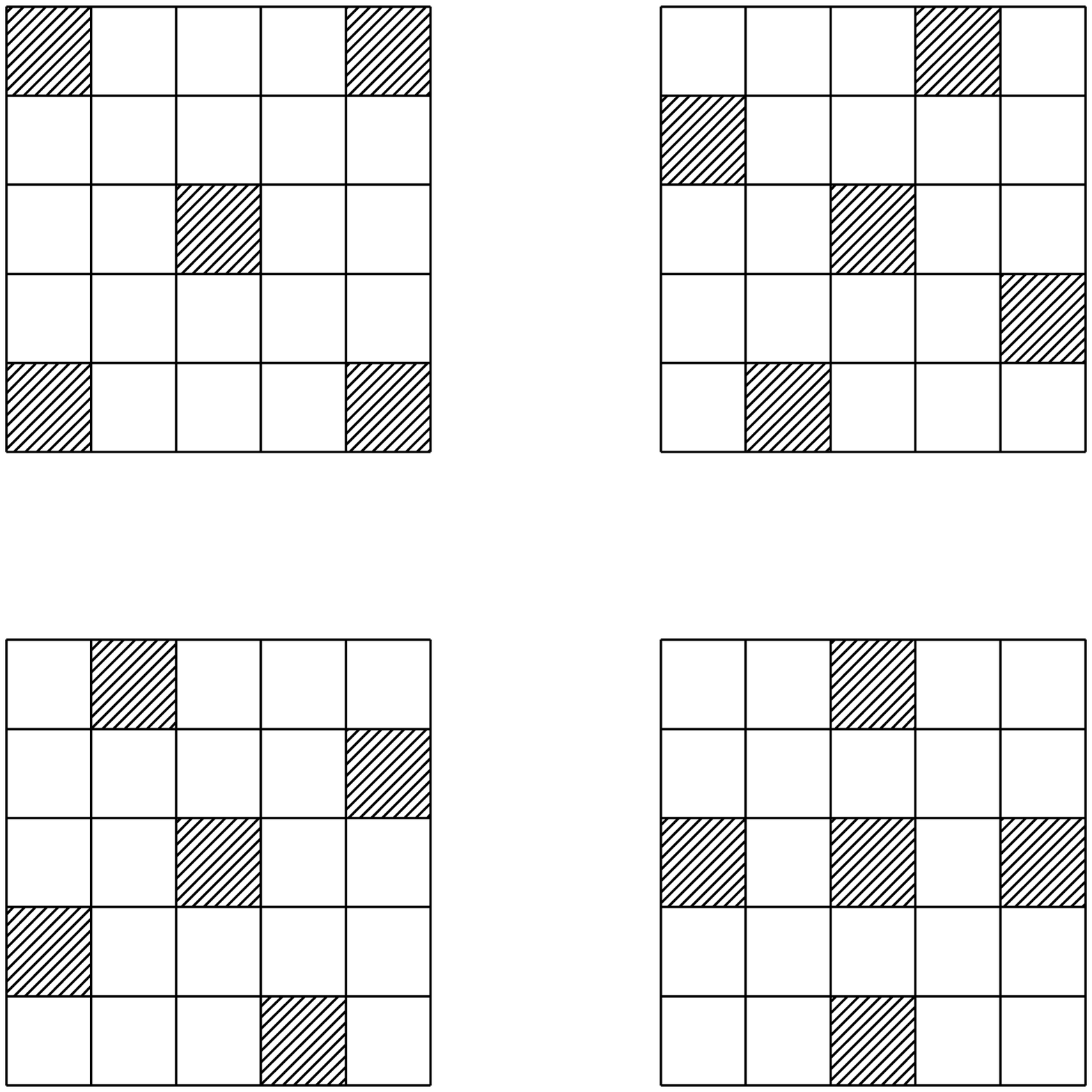}
\caption{  
 The Patterns 0 to 3 used to subdivide
a five by five square.}
\label {fig1}
\end{figure}

We illustrate all these phenomena for a set of dimension 1 in 2-space.
We then generalize to the case of a set of dimension 2 in 3-space.
Our example is constructed as follows:
Divide the unit square into 25
equal small squares and fill the central square and the four
corners. We call this ``Pattern 0.'' Similarly, we can fill 5 among
25 squares by Patterns 1, 2, and 3 as shown in Fig.~\ref{fig1}

\begin{figure}[ht]
\epsfxsize=6cm
\epsffile{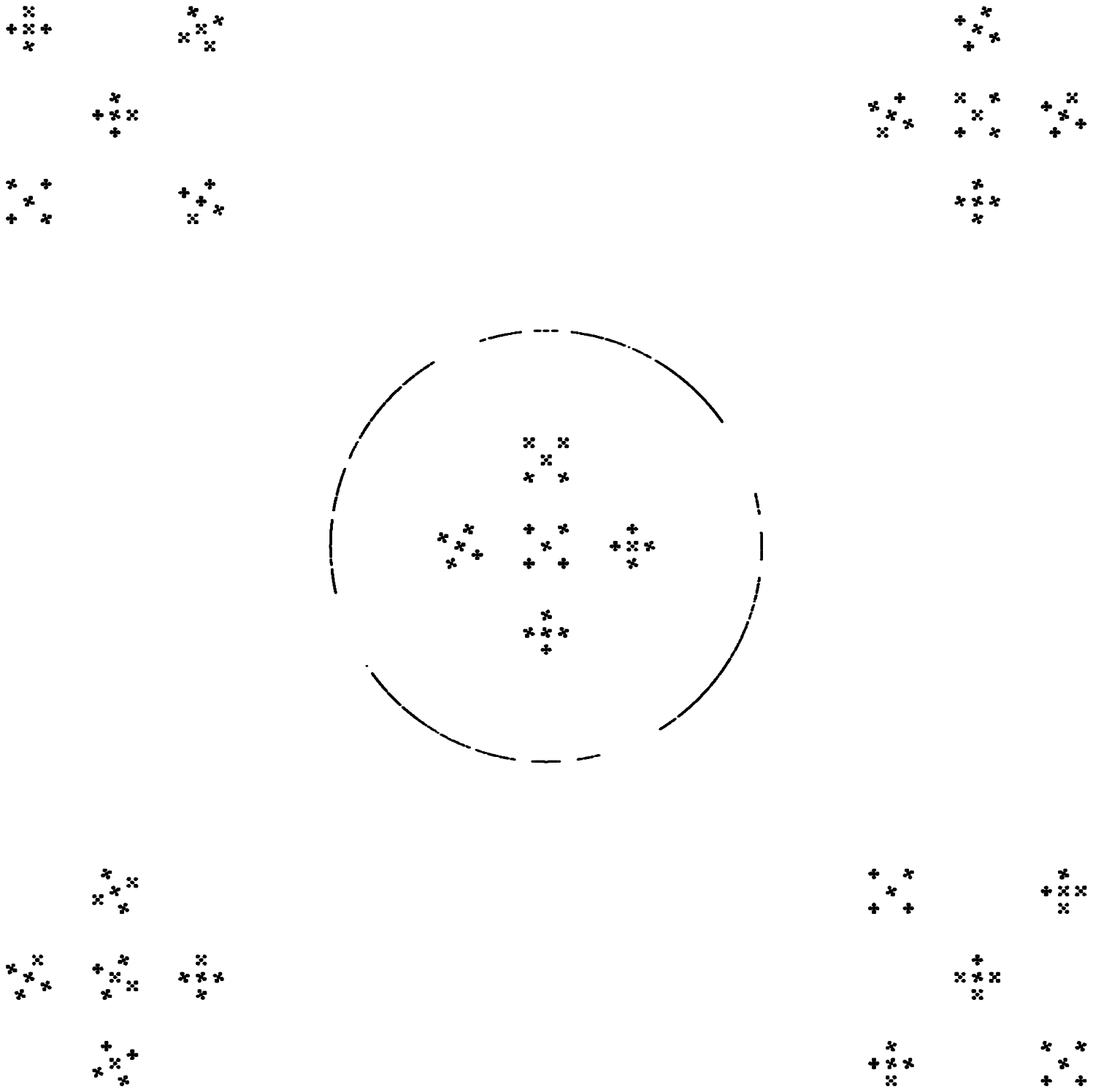}
\caption{A fractal of dimension 1 in 2-space, and its projection onto
a circle. There are 7 levels of recursion.}
\label{fig2}
\end{figure}

The
fractal of Fig.~\ref{fig2} is then obtained by dividing the square recursively,
choosing at
each level randomly one of the  4 patterns.
(An even more homogeneous projection is obtained
by choosing at level $n$ the pattern  $(n\,{\rm mod}\,4)$
of Fig.~\ref{fig1}.)
These fractals have dimension 1, finite Hausdorff measure $H^1$, and are
of the singular type described above.
Of course, space gets quite empty far away from the origin, but
such fractals ``block'' the skylight in
almost all directions for an observer at the center, and hence she will
see an almost uniformly black sky as in Fig.~\ref{fig2}. 
If the observer is not at
a center, but still on a ``galaxy,'' as we are, this argument
continues to 
hold after enough iterations.
To understand why the lacunarity decreases, we scale each point from
the center to a fixed distance: $(x,y)\mapsto (x/r,y/r)$, where
$r=(x^2+y^2)^{1/2}$. In Fig.~\ref{fig3}, we blow each ``inner'' point up to a
position on the outer set of squares.

Since there is always one of 4 squares colored in each of the Patterns
0--3, we see that the apparent open space can be at most about
$3/5^{\rm th}$ of what it was at the previous level. Thus we get an
estimate that {\em the maximal angular void scales like $(3/5)^n$ as
the number $n$ of levels grows.} This answers one of Peebles'
questions: There need not be any sizeable voids
in the projection of a set
of dimension 1 in 2-space (or for that matter, a set of dimension 2 in
3-space).

We next discuss how the projection onto the celestial globe can vary,
depending on whether we show apparent size (as in Fig.~\ref{fig3}) or just a
pixel (as in the circle around Fig.~\ref{fig4}).
To simplify the discussion,  we assume that all 
galaxies are small spheres of fixed diameter $\epsilon$. 
The projection of a galaxy 
at distance $r$ has apparent size  $\sim \epsilon / r$.
One can view this in one of two ways: Either the remote galaxies
have very small projections (somehow less than a pixel), or the
close-by ones have very large projections. Note that we discuss here
for simplicity a universe of galaxies of {\em equal} size. How big 
is the area covered by these projections?
Since 
\begin{equation}
H^1({\bf A}) \,=\, \lim_{\delta \rightarrow 0} \sum_i {\rm diam}(S_{i,
\delta})  ~,
\end{equation}
and there are $\sim C$ galaxies at a distance $r$ and 
${\rm diam}(S_{i, \delta}) = \delta$, we find that the number of
galaxies ($\propto$ area of pixel projection) in a
ring extending from $r_{\min}$ to $r_{\max}$ is given by
\begin{equation}
N(r_{\min},r_{\max}) \,\sim\, \int_{r_{\min}}^{r_{\max}}dr'\, C  \,\sim\, C
(r_{\max}- r_{\min})~,
\label{counting}
\end{equation}
while the {\em projected area} is
\begin{equation}
\int_{r_{\min}}^{r_{\max}} dr'\, C/r' = 
\,\, C\log\left(r_{\max}/ r_{\min}\right)~.
\label{area}
\end{equation} 
(The area is smaller when the projected galaxies start to
overlap, {\em
i.e.,} for very fat rings.)
These equations explain why projected galaxies drawn as points 
(Eq.~\ref{counting})
look more homogeneous after increasing the size of the annulus
$(r_{\max}, r_{\min})\rightarrow \lambda(r_{\max},r_{\min})$; whereas a
projection of the apparent area occupied by galaxies (Eq.~\ref{area}) is invariant under
scaling transformations. The homogenization resulting in a point
projection (Eq.~\ref{counting}) has been used as proof for the
homogeneity of the universe\cite{Davis}. However, in our simple
fractal model, homogenization occurs as well, for shells with
$r_{\max} > 5^4 r_{\min}$, and thus the reasoning in \cite{Davis} is
not conclusive. At small radii, finite size effects dominate, as in
Fig.~1 of \cite{Davis}. Using smaller sub-divisions in the construction of
the example,
one can lower the necessary shell-width to $r_{\max}\sim 2 r_{\min}$.

In the area projection, most of the area is covered by close-by disks, 
because they are larger. On the other hand, Figure 
3.10 (of Radio-galaxies) in  Peebles' book \cite{Peebles} shows all 
the disks of the same size 
(namely as small pixels, which correspond to the projected measure of
a galaxy at distance about $1/r_{\max} \sim \delta$). 

To illustrate this point, 
we construct a 2-dimensional set in 3-space, using an adapted
variant of the algorithm described above. (We take a $3\times 3\times
3$ cube and select 9 among the 27 possible little cubes, producing
essentially 3 patterns with the central cube and 8 other cubes
occupied. This produces higher density than $5^3$.)
In
Fig.~\ref{fig3d1} we show the apparent size and the pixel
projection of
this set onto the 2-dimensional celestial globe using the
Aitoff-projection. One clearly notices that the pixel projection looks
less ``lacunary'' than the apparent size projection.
In Fig.~\ref{fig3d2} we show the same projection when the galaxies are
``opaque.'' This reduces the local maxima of the apparent density of
galaxies. Clearly, the projection of our 3D fractals are by far less
isotropic than the observed galaxy distribution. Whether this is due
to the fact that the effects discussed in this paper are not relevant
to the galaxy distribution or whether it is due to our rather
naive algorithm is not clear. A  more detailed study of this is left 
as a future project.
But this is not our main point here. We just want to demonstrate that
from an isotropic angular projection one cannot conclude that the
3-dimensional galaxy distribution is homogeneous on any scale.

{\bf Discussion}: We have analyzed voids and projected points of a
2 dimensional fractal set in 3-space projected onto the celestial sphere.
First, we have shown that the voids in the projection
can be very small, even if they are large in the original space distribution.
We also have
shown that the apparent fluctuations of the
projection of a set of ``galaxies'' depend on whether they are
projected with fixed or apparent size. In the first case, the
projection is more uniform than in the second. This explains in part
the apparent contradiction between the relatively regular projection
of a 2-dimensional set in 3-space and its intrinsic fractal nature.
Of course, this simple geometric analysis does not answer the much
deeper apparent cosmological contradictions between the isotropy of
the universe and the  fractal dimension of 2 of the galaxy
distribution as found in \cite{Pietronero}. It seems to us that this 
field is still wide open to speculation. Finally, if we
assume that there is an observational limit to how close to each
other two galaxies can be observed (by drawing a little ``opaque'' disks of
equal size around each observed galaxy), the pixel projection
such as Fig.~\ref{fig3d1} above become somewhat less
fluctuating, see Fig.~\ref{fig3d2}. The study of this, and similar 
effects in more detail will have to await further work.
	
\smallskip
\noindent{\bf Acknowledgments}. We thank P.~Mattila for very useful
correspondence concerning fractal sets, and D. Pfenniger for remarks. 
L.P. thanks B. Mandelbrot for helpful discussions.
This work was partially supported by the Fonds
National Suisse.

\newpage


\begin{figure}[ht]
\vspace{0.5cm}
{\Large Fig. 3}
\vspace{0.5cm}
\caption{Five levels of the same fractal as in Fig.2. 
Each ``inner''
feature is scaled outwards as shown by the lines for one
feature. 
The corners contain the same feature as the central square.
The outermost features of the central square
are mapped as indicated by the radial lines.
Going deeper into the recursion, each
successive feature is projected clockwise
to the boundary until
one reaches again the corner squares in light gray which
are mappings of the innermost feature.}
\label{fig3}
\end{figure}
 

\begin{figure}[ht]
\vspace{0.5cm}
{\Large Fig. 4}
\vspace{0.5cm}
 \caption{The same as Fig.3, but now with ``opaque''
galaxies. Note that the projected density on the unit circle becomes
quite uniform.} 
\label{fig4}
\end{figure}


\begin{figure}[ht]
\vspace{0.5cm}
{\Large Fig. 5a}
\vspace{0.5cm}
\end{figure}

\begin{figure}[ht]
\vspace{0.5cm}
{\Large Fig. 5b}
\vspace{0.5cm}
\caption{The projection of a 2-dimensional set ($\sim 30000$
galaxies) onto the celestial
globe. The calculation involves 9 levels.
Top: Galaxies shown at apparent size. Bottom: Galaxies shown
at equal size (pixel projection).}
\label{fig3d1}
\end{figure}


\begin{figure}[ht]
\vspace{0.5cm}
{\Large Fig. 6}
\vspace{0.5cm}
\caption{The projection of a 2-dimensional set onto the celestial
globe. Galaxies are shown
equal size and are opaque.}
\label{fig3d2}
\end{figure}

\end{document}